\documentclass[preprint]{aastex}

\def\gsim{\;\rlap{\lower 2.5pt
 \hbox{$\sim$}}\raise 1.5pt\hbox{$>$}\;}
\def\lsim{\;\rlap{\lower 2.5pt
   \hbox{$\sim$}}\raise 1.5pt\hbox{$<$}\;}

\slugcomment{}

\shorttitle{Long Period Exoplanets}
\shortauthors{Scharf \& Menou}

\begin{document}


\title{Long period exoplanets from dynamical relaxation}


\author{Caleb Scharf\altaffilmark{1}}
\affil{Columbia Astrophysics Laboratory, Columbia University, 550 West 120th St., MC 5247, New York, NY 10027}\email{caleb@astro.columbia.edu}

\author{Kristen Menou}
\affil{Department of Astronomy, Columbia University, }
\email{kristen@astro.columbia.edu}


\altaffiltext{1}{Columbia Astrobiology Center}


\begin{abstract}
Recent imaging campaigns indicate the likely existence of massive
planets ($\sim 1-10$ M$_{J}$) on $\gsim 1000$ year orbits about a few percent of 
stars. Such objects are not easily explained in most current planet
formation models. In this Letter we use ensembles of 100 N-body
simulations to evaluate the potential for planet scattering during
relaxation of dynamically active systems to explain the population of giant planets with
projected separations up to a few 100 AU. We find that
such a mechanism could indeed be at play, and that statistical samples of
long period planets could place interesting constraints on early stage
planet formation scenarios. Results from direct imaging and
microlensing surveys are complementary probes of this dynamical
relaxation process.
\end{abstract}


\section{Introduction}

The ongoing discovery of exoplanets is providing a wealth of
opportunities to test models of planetary system formation, and has
already revealed that our own solar system is not necessarily typical
(see \citet{udry2007} for a review). Due to the natural biases of 
radial velocity and transit
detection methods a major focus has been on short period planets, of all
ages. Direct imaging efforts are biased oppositely, towards the
detection of long period, young, and massive planets \citep[see,
e.g.,][for a review]{beuzit2007}. Microlensing searches are also
sensitive to longer period objects, independent of system age
\citep[see, e.g.,][for a review]{gaudi2008}. Recent direct imaging
efforts have yielded constraints on the likelihood of certain
classes of long period gas-giant planets
\citep{masciadri2005,biller2007,kasper2007,nielsen2008}, together with
a number of increasingly robust claims of detection of long period
planets, or sub-stellar companions \citep[e.g.,][]{neuhauser2005,mohanty2007,oppen08, laf08}; and in particular \citet{kalas2008} and \citet{marois2008}.
These objects have projected separations of $\sim 20$ to a few 100 AU.
The result of \citet{laf08} (LJvK08) is interesting, since it
claims detection of an $8^{+4}_{-1}$ M$_J$ object at apparent separation
of 330 AU from the parent star (1RXS J160929.1-210524) - an
approximately solar-mass star in the young ($\sim 5$ Myr) Upper
Scorpius association. Taken at face value this
represents one positive detection out of $\sim 80$ potential planet
hosting stars, or a detection rate of $\sim 1$\% - although clearly
this is a very crude evaluation since the detectability of companions
is a complex function varying from star to star. 

Most protostellar/proto-planetary disks are approximately 100 AU in
radius, and core accretion models predict giant planet formation close
to the water-ice snowline ($< 10$ AU), where coalescence timescales
are close to, or within, the disk lifetime against photoevaporation
(e.g. \citet{hollenbach2000}). Gravitational instability models
\citep{boss1997} might have a higher probability of forming massive
planets at large radii - but this would require quite unusually
massive and extended disks. As pointed out by LJvK08, this suggests
that migration and/or scattering mechanisms might be necessary to
explain a planet at projected separation of 330 AU. Another
possibility is that highly mass-imbalanced wide binary systems could
form, with stellar and sub-stellar companions.

In this Letter we present the results of numerical simulations of
ensembles of 100 planetary systems that capture the effect of strong
dynamical evolution through planet-planet scattering. We present a
number of proof-of-principle predictions for objects with large
semi-major orbital axes ($>100$ AU), in particular those in young
systems. Our methodology stems primarily from the work of
\citet{juric08}, and other investigations of dynamical effects
\citep[e.g.,][]{ papaloizou2001, adams2003, ida2004, veras2004, adams2006, debes2006,
chatterjee2008, ford2008, raymond2008, thommes2008b}. \citet{juric08}
suggest that, for the purposes of studying dynamical evolution, planet
formation may be simplified into two stages. The initial
episode (Stage 1) involves the assembly and evolution of a
circumstellar and proto-planetary disk over $\sim 10^6-10^7$
years. During Stage 1, planetary embryos are formed, destroyed, and
interact strongly with their disk surroundings. Stage 2 takes place
once the major part of the gaseous proto-planetary disk has
dissipated. It involves the dynamical evolution of any planets that
remain, including collisions, mergers, scattering, and significant
planet ejection from the systems over the following tens to hundreds
of millions of years. Using large ensembles of statistically identical
initial conditions for planetary systems \citet{juric08} demonstrated
that, in quite general circumstances, aspects of the observed
distribution of {\em giant} exoplanet orbital eccentricities can be
reproduced.  This result is not strongly dependent on initial
conditions as long as systems are dynamically active, or partially
active. It also demonstrates the utility of studying the {\em
statistical dynamical} properties of a population of planetary
systems, rather than trying to identify a physical process that
prescribes orbital characteristics in any single system.

Planets in dynamically active systems experience strong mutual
interactions and frequent encounters and undergo relaxation. This
typically occurs when the spacing relative to the mutual Hill Radii \citep[e.g.,][]{henon1986} is small.
Partially active systems contain only a subset of objects undergoing
strong and frequent interactions, and while {\em inactive} systems
still evolve, there are typically few or no major encounters in this
case. Both active and partially active systems tend to lose
planets as a function of time (instability timescales are given by \citet{chatterjee2008}).
 This occurs through collisions (and
mergers), capture by the parent star, or very often through complete
ejection from the system. Thus, the Stage 2 dynamical paradigm
predicts {\em both} a large population of ejected planets and a
systematic excess of planets in very young systems compared to older
systems.

\section{Simulations and methodology}
We employ the publicly available N-body integration routines in the
MERCURY6 package \citep{chambers1999} to follow the dynamical
evolution of statistically identical ensembles of planetary
systems. MERCURY6 also allows for close encounters and fully inelastic
(no fragmentation) star-planet or planet-planet collisional
mergers. We have found that a high-accuracy Burlisch-Stoer integrator
typically provides relative energy errors of less than $10^{-5}$ up to
integration times of $10^7$ years for the systems we are modeling. For
simplicity we therefore use the Burlisch-Stoer BS2 integrator in
MERCURY6 and restrict our integrations to $10^7$ years - commensurate
with the ages of the young planetary systems we hope to place
constraints upon. Our methodology is otherwise similar to
that of \citet{juric08}.  We note that for dynamically active systems
all but the final $\sim 10$\% of planet ejection/loss events have
occurred by $10^7$ years.  Our simulations  do not include the
effects of tidal dissipation between planetary objects and the parent
star.  Dissipation due to star-planet tides can act to dampen orbital
eccentricities, and is an important physical process for planets with
short periods.  We choose to circumvent these issues by ignoring
planets with semi-major axes of less than 0.1 AU in our final
results. This is well motivated in that tidal damping is expected to
operate on timescales ranging from $\sim 10^{9}$ to $\sim 10^{12}$
years for objects with $\sim 20$ day orbital periods around solar mass
stars \citep[e.g.,][and references therein]{matsumura2008}.

We adopt a (somewhat arbitrary) physical ejection radius of 2000 AU, but note that for semi-major axes beyond $\sim1000$ AU, secular effects from Galactic tides may become important \citep[see,
e.g.,][]{higuchi2007}. It is possible that these tides would cause
some of the planets we label as ejected to remain on bound orbits at
very large separations. We neglect this possibility altogether in the
present study.

\subsection{Initial populations of objects}

We perform two sets of runs, each with statistical ensembles of 100
systems. We populate the systems with 10 planets and
adopt initial properties such that the systems are dynamically active.
Masses are drawn randomly from a uniform distribution in $\log(M)$ and
are limited to a range of 0.1-10 M$_J$. Initial orbital
inclinations ($i$) for all runs are drawn randomly from a Rayleigh
(Schwarschild) distribution - where the peak, or mode, is equal to the
distribution width parameter $\sigma$, and we set $\sigma_i=0.3$
degrees. Orbital eccentricities are also drawn from a Rayleigh
distribution with $\sigma_e=0.3$. In one run, that we
denote as $R_{30}$, the initial semi-major axes are drawn randomly
from a uniform distribution in $\log(distance)$ between 0.1 AU and 30
AU. In the second run, denoted as $R_{100}$, the outer truncation is increased to 100 AU.  All systems contain a
central stellar mass of $1$M$_{\odot}$.

The $R_{30}$ and $R_{100}$ runs illustrate as simply as possible the dependence
on initial conditions of the dynamical relaxation process. The
two leading scenarios for planet formation, core accretion and direct
gravitational collapse, should lead to significantly different initial
orbital outcomes for young planets
\citep[e.g.,][]{lissauer1993,boss1997,papterquem2006}. Each of these
scenarios faces challenges 
\citep[e.g.,][]{matzlevin2005,rafikov2005,dong2008} and they could in
principle both operate in individual systems or in the proto-planetary
population as a whole \citep[e.g.,][]{ribasmiralda2007}. Given these
complications, what is important in the present context is that
the late dynamical architecture of exoplanetary systems could 
potentially allow us to discriminate between competing formation
scenarios. Specifically, formation via core accretion is favored in
the vicinity of the snow-line in the proto-planetary disk,
typically well within a few tens of AU, while gravitational collapse
is favored at larger orbital distances. This is the primary motivation
for the two types of runs $R_{30}$ and $R_{100}$ performed here.


\subsection{Monte Carlo ``snapshots"}

In order to model apparent separations of
planet-star systems from our simulation outputs we must take into
account the orbital characteristics and potential orientation of
systems relative to a hypothetical observer. We adopt a simple
algorithmic approach. First, the semi-major axis, eccentricity, and
orbital period are used to construct a probability distribution for
the instantaneous radial star-planet distance, assuming an observation
made at a random time during the orbital period. We then assume that
the orbital plane is randomly orientated with respect to the observer
and project the star-planet radial separation onto the sky to obtain
an apparent separation.

We can draw an arbitrary number of ``snapshot" projections for all
remaining bound planets in our simulations.  Since this step is not
computationally intensive, the actual number of snapshots is simply
determined by the need to adequately populate the mass -- projected
separation plane. Here we generate 1000 such
projections per individual planet, irrespective of any other bound
planet in the system. For simplicity, we then generate a cumulative
distribution of projected separations for all individual planets,
ignoring entirely conditional probabilities associated with multiple
planetary systems. This is sufficient for our present purpose, focused
on the apparently most distant planet, but a more careful analysis
would clearly require conditional probabilities for projected
separations to be accounted for.

Uncertainties in the separation distributions are estimated
from the dominant Poisson statistics of the original simulations with
 a total of only $10 \times 100$ statistically-identical planets, as
opposed to the more numerous projection realizations. For instance, if
$N$ uniquely tagged planets, out of the total $1000$ original ones,
contribute to the statistics beyond some fixed projected separation,
an uncertainty $\sqrt{N}/N$ is applied to the apparent separation
distribution.  Plotted error bars/envelopes are $1\sigma$.

\section{Results}
In Figure 1 we plot the mean number of  planets per system as
a function of time for the two runs $R_{30}$ and $R_{100}$. The more
closely packed planets in $R_{30}$ result in a faster decay, and
by $10^7$ years, when both runs are much more dynamically relaxed, the
mean number of remaining planets in $R_{30}$ is slightly lower than
that in $R_{100}$, with $\sim 2$ planets per system in both
cases. These results are in good agreement with those of
\citet{juric08}, who demonstrated that integrating to $10^8$ years
(by employing a symplectic integrator) results in only moderate
further evolution, owing to the loss already of an average 80\% of the
initial population by $10^7$ years.
 
By $10^7$ years in run $R_{30}$, 25\% of the initial population have
collided with the central star, 12\% undergo collision and merger with
other planets and 44\% are ejected (the majority by scattering into an unbound $e>1$ orbit, some by
passing beyond 2000 AU, see below). For $R_{100}$ 21\% collide with
the central star, 11\% undergo collision and merger, and 44\% are
ejected. This indicates that the primary difference in final planet
numbers at $10^7$ years between $R_{30}$ and $R_{100}$ is almost
entirely due to the higher rate of collision with the central star in
$R_{30}$. 
 
In Figure 2 the eccentricity versus semi-major axis is plotted for
all objects at several timeslices for the $R_{100}$ run to illustrate  the orbital evolution of the population. 
In the first $\sim 10^6$ years a significant number of objects are scattered onto high eccentricity and large semi-major axis orbits. The great majority ($>99.2$\%) of objects that reach eccentricities $e>0.9$ become unbound ($e>1$) due to scattering close to periastron.

In Figure 3 we plot the cumulative number of planets beyond a given
apparent separation. In both $R_{30}$ and $R_{100}$ runs, some
outward diffusion of the planet population is clearly seen by
comparison to the initial distribution.  Cumulative distributions
after $10^6$ and $10^7$ years are shown, and only modest evolution is
seen at these late times. In both runs, there is some reduction in the
fraction of objects between a few AU and $\sim 100$ AU, from $10^6$ to
$10^7$ years. 

For comparison with the putative planet detection of LJvK08, we find
that the population of planets with 330 AU apparent separation or
greater in the $R_{100}$ run is $0.9(\pm{0.6})$\% with a variation of
only $\sim 0.2\%$ between $10^6$ and $10^7$ years. In the $R_{30}$
run, the corresponding numbers are $0.6(\pm{0.3})$\% at $10^6$ years
and $0.2( \pm {0.2})$\% at $10^7$ years.  We note that these results are broadly consistent with
\citet{veras2008} who find that $\sim 0.5$\% of dynamically evolved (scattered) systems at late times ($>10^7$ years) could still harbor  a long-term stable planet with $e>0.8$ and $a> 1000$ AU.

For a minimum apparent separation of 100 AU, at $10^6$ years, these
fractions increase to $7.1(\pm 2.5)$\% in the $R_{100}$ runs and
$2.3(\pm1.0)$\% in the $R_{30}$ runs. Later, at $10^7$ years, there is
little evolution in the $R_{100}$ results, but a drop to $1.4(\pm
0.6)$\% in the $R_{30}$ results.

Although these numbers are not suitable for a direct comparison to
current imaging survey constraints, we nonetheless note that the
constraint of \citet{nielsen2008} that less than 20\% of systems
should harbor $>4$M$_J$ planets between 20 and 100 AU is broadly
consistent with the results of our simulations. The $R_{30}$ runs
indicate that less than $\sim 16$\% of planets of all masses would be
expected at larger than 20 AU apparent separations (at $10^7$ years).
The corresponding $R_{100}$ result is less than $\sim 30$\%.

The results shown in Figure 3 do not discriminate between planets of
different masses as a function of projected separation. This is a
particularly important issue for direct imaging surveys, with yields
depending on a combination of ages, masses and projected
separations. We believe that a larger number of statistical
realizations than performed here is needed to address this point
reliably and postpone a more detailed investigation to future
work.

\section{Discussion}

Dynamical relaxation of young planetary systems can produce a
population of long period planets within $10^6-10^7$ years of planet
formation.

The results presented here are subject to a number of
limitations. Although the outcome of dynamical relaxation of active
systems in terms of eccentricities is not strongly dependent on the initial conditions set during
Stage 1 planet formation \citep{juric08}, the precise relationship between other population characteristics
at $\sim 10^7$ years in Stage 2 and initial parameters needs further
exploration.  For example, the differences between our $R_{30}$ and
$R_{100}$ runs indicate that information on semi-major
axis distributions is indeed retained in the population
of objects at later times.  The population of long period planets could therefore 
potentially provide important constraints on the
outcome of Stage 1 planet formation.

Furthermore, to make meaningful comparisons with 
observational surveys, one must allow for at least two biases: 1) Not
all stars surveyed will necessarily form a population of giant planets
during Stage 1. 2) Not all systems with giant planets will be
dynamically active at the start of Stage 2. For instance,
\citet{juric08} estimate on the basis of the observed distribution of
exoplanet eccentricities that perhaps 75\% of currently known
planetary systems have been dynamically active. Less dynamically
active systems, which were not directly addressed here, may provide
better records of the planet formation outcome but at the expense of a
smaller range of projected separations (with respect to the
dynamically relaxed apparent separations shown in Fig.~3).

The detailed strategy of an imaging survey is also important for
making meaningful comparisons with theoretical expectations. If the
planet formation phase (Stage 1) lasts between $10^6$ and $10^7$
years, the timescale for disk dispersal/evaporation
\citep[e.g.,][]{hollenbach2000}, then a ``blind" survey of young stars
- where the target objects span a wide range of ages up to a few
$10^7$ years - may not be well suited to probe the earliest Stage 2
phase. Indeed, the fast (Stage 2) relaxation phase  (Figure 1),
relative to a longer duration Stage 1 gaseous phase, suggests that it
may be statistically difficult to identify the systematically larger
number of planets expected in younger, dynamically-active
proto-planetary systems. On the other hand, \citet{thommes2008a}
present evolutionary scenarios based on mean-motion resonance locking
and remnant planetisimal disk evolution in which the transition from
Stage 1 to the Stage 2 relaxation occur much more gradually. Finding a
 larger number of planets in younger systems would provide strong
observational evidence in support of the dynamical relaxation scenario.


We conclude that the detection of  massive planets at 
apparent separations of a few 100 AU need not be a challenge to the core accretion model of
planet formation if dynamical relaxation occurs through extensive
planet-planet scattering. Furthermore, the number of very long period
planets resulting from scattering during the first few Myrs following
disk dispersal may be consistent with current upper limits, and
claimed detections, from all existing planet imaging campaigns.

These are encouraging but very preliminary results. They suggest that
a more systematic exploration of dynamical relaxation outcomes for
long-period planets, as a function of initial conditions and system
age, could provide a valuable additional means to interpret the existing
and future yields of direct imaging surveys. Furthermore, since microlensing
surveys probe the same projected planetary separations,
but typically at much later system ages, they provide
very useful complimentary information on the late stages of the
dynamical relaxation process. Together such data would allow
consistency checks to be performed on the systematic trends expected
between early and late stages of orbital evolution in this promising model.




\acknowledgments
The referee is thanked for comments that helped improve the paper. CS acknowledges the support
of Columbia University's Initiatives in Science and Engineering, and NASA Astrobiology: Exobiology grant \# NNG05GO79G.

\clearpage



\begin{figure}
\plotone{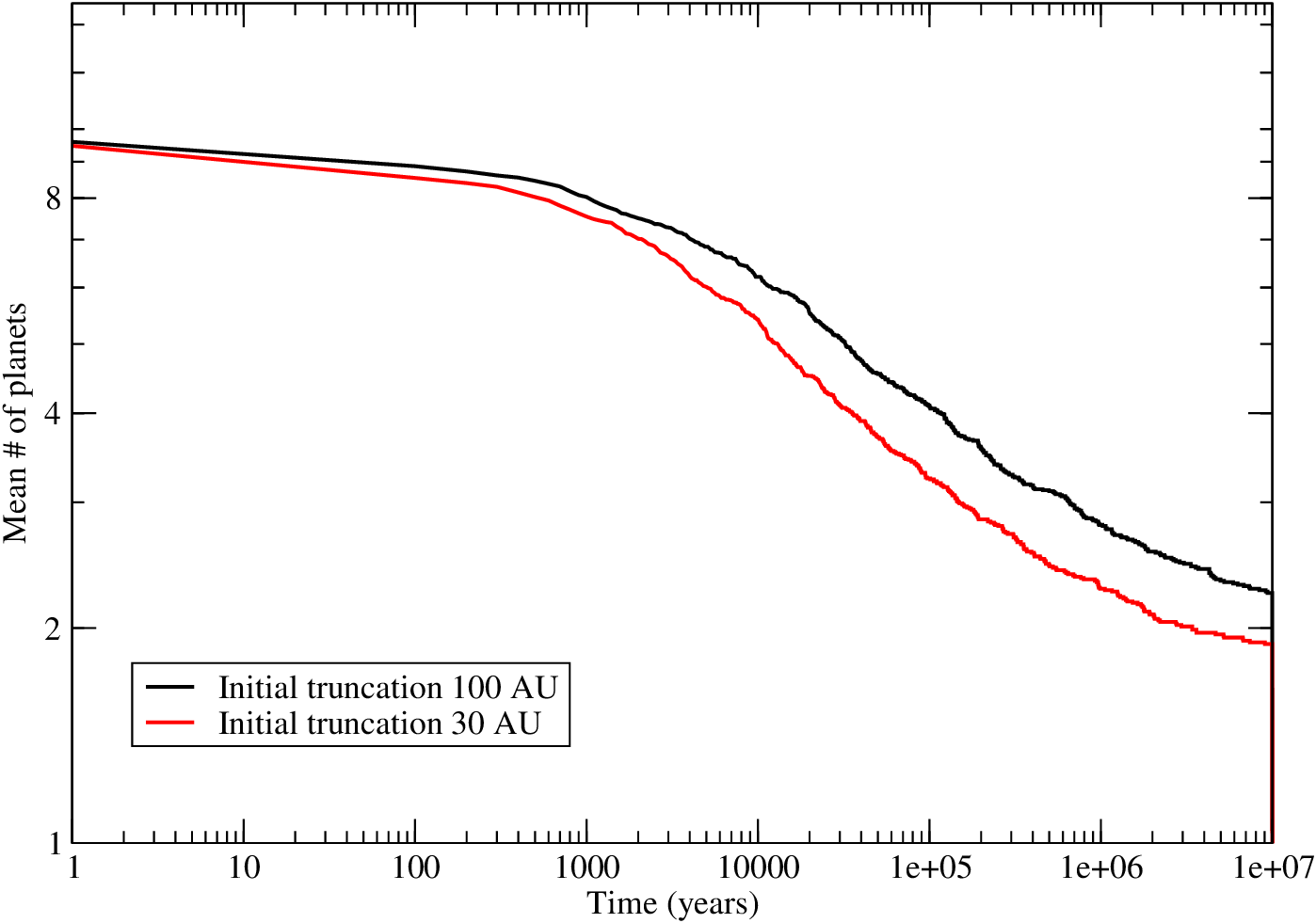}
\caption{The mean number of remaining planets per system is plotted as
a function of time for runs $R_{30}$ and
$R_{100}$. Most collisions (planet-planet and planet-star) and
ejections occur within the first $10^5-10^6$ simulated
years.\label{fig1}}
\end{figure}

\clearpage

\begin{figure}
\plotone{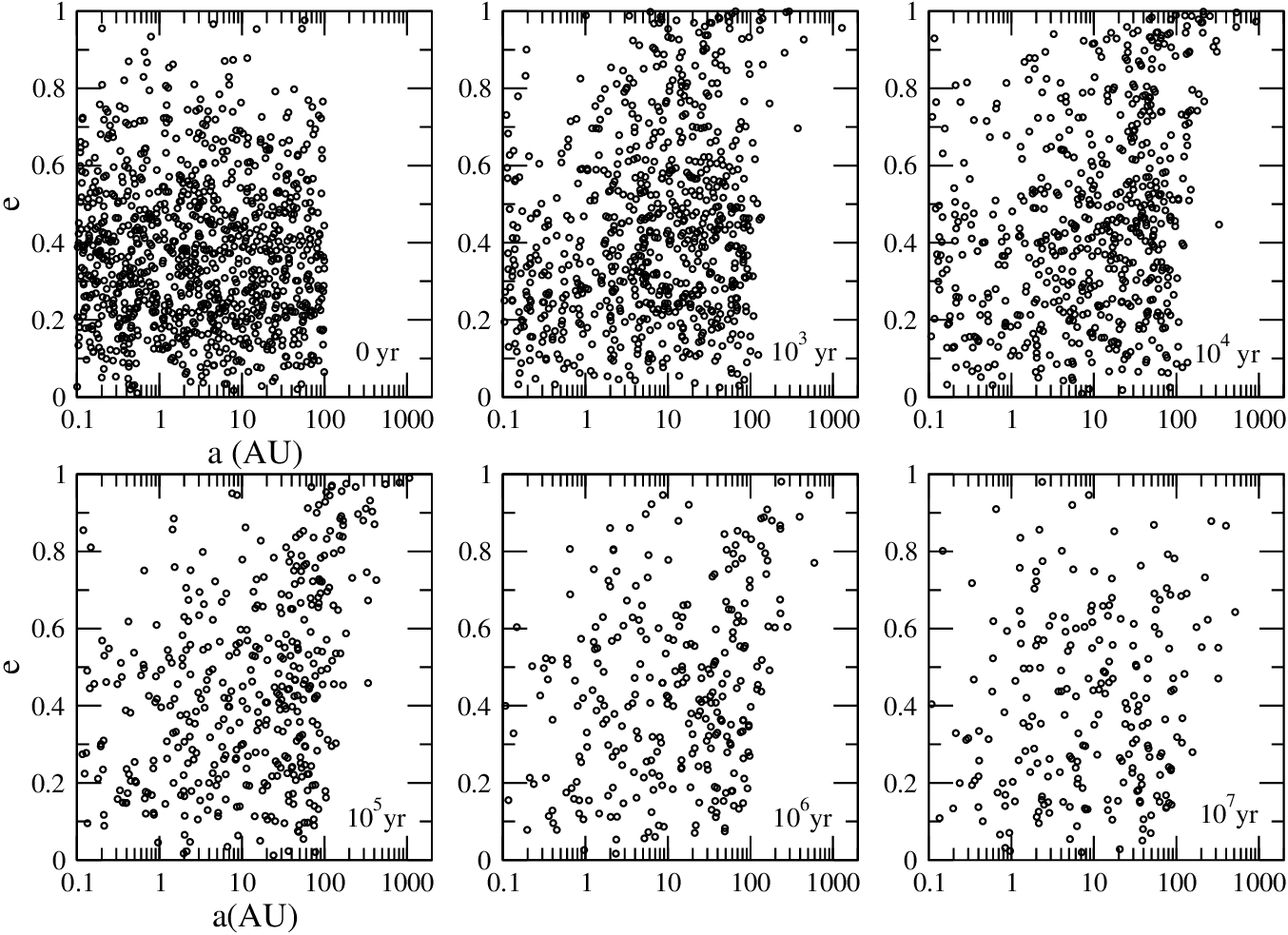}
\caption{The orbital eccentricities and semi-major axes of objects plotted at a number of times for the $R_{100}$ run. Starting at the upper left and moving right, plots show the distribution at 0, $10^3$, and $10^4$ years. Lower panels, left to right, are for $10^5$, $10^6$ and $10^7$ years. Of the highest eccentricity objects ($e>0.9$) at any given time, less than $\sim 0.8$\% remain on bound orbits by $10^7$ years.
\label{fig2}}
\end{figure}

\clearpage

\begin{figure}
\plotone{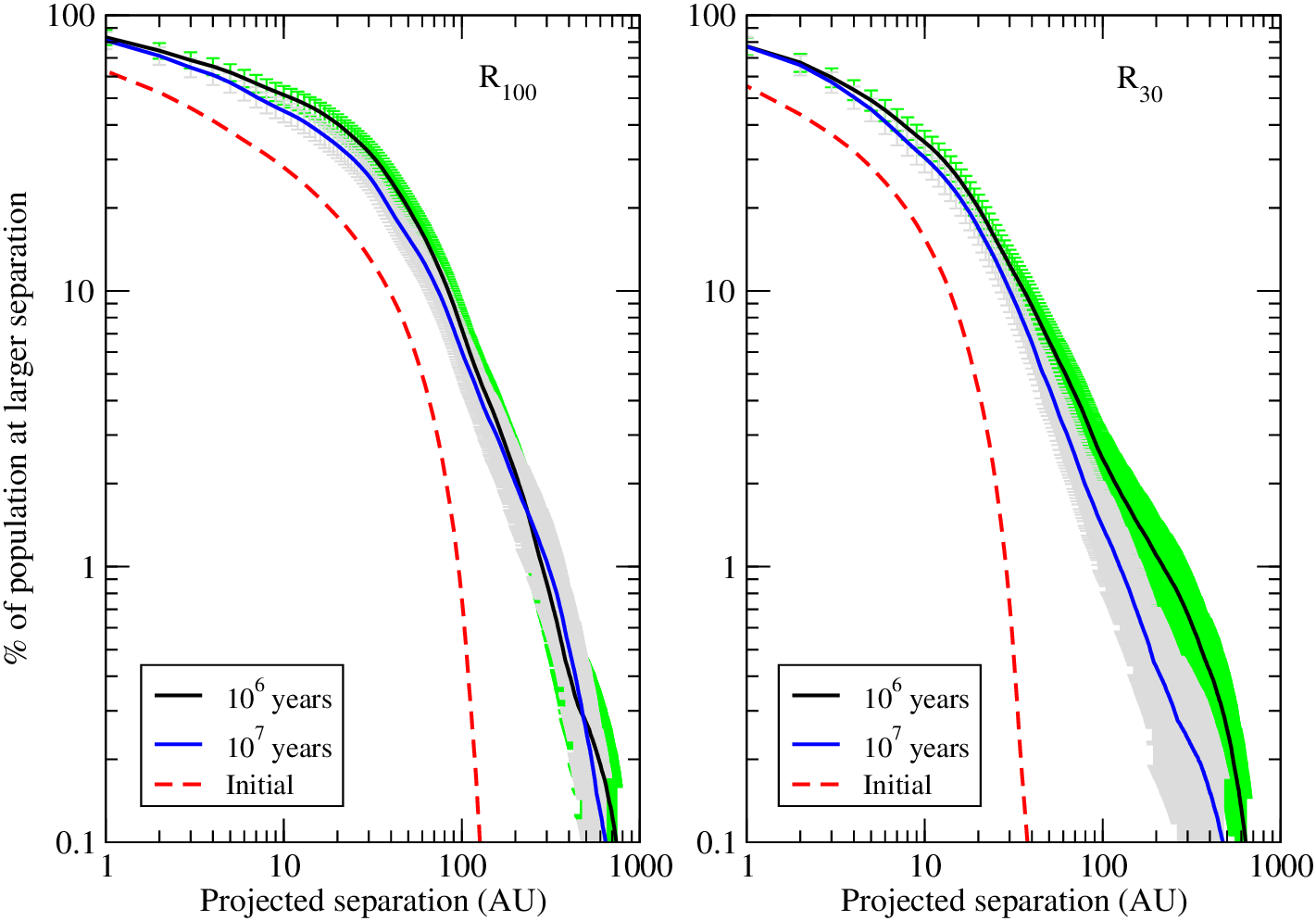}
\caption{The percentage of the remaining planet population with
apparent separation greater than a given value is shown for the two
models explored. {\it Left panel:} $R_{100}$ ensemble results. Black
curve and green error bars correspond to the population after $10^6$
years, blue curve and grey error bars correspond to the population
after $10^7$ years.  Errors are computed from the Poisson uncertainty
in the underlying number of planets used to seed the Monte Carlo
projected separation distributions. The red dashed curve shows the
initial conditions.  {\it Right panel:} $R_{30}$ ensemble results. In
both panels, the consequences of an outward diffusion of planets from
their initial distribution are clearly seen. Planets in the tail of the
projected separation distribution typically have eccentricities $e
\gsim 0.4$ (see Figure 2).
\label{fig3}}
\end{figure}

\clearpage

\end{document}